\documentclass[runningheads]{cl2emult}

\usepackage{makeidx}  
\usepackage{graphicx} 
\usepackage{subeqnar} 
\usepackage{multicol} 
\usepackage{physbb}   
\makeindex            

\newcommand{\element}[2][]{$^{#1}$#2}

\begin{document}

\title*{The Hubble Constant from Type Ia Supernovae in Early-Type
  Galaxies \footnote{To be published in: Kundt W., van~de~Bruck C.
    (Eds.) Cosmology and Astrophysics: A collection of critical
    thoughts.  Lecture Notes in Physics, Springer}}

\toctitle{The Hubble Constant from Type Ia Supernovae in Early-Type
  Galaxies}

\titlerunning{The Hubble Constant from Type Ia Supernovae in
  Early-Type Galaxies}

\author{Tom Richtler\inst{1} \and Georg Drenkhahn\inst{2,1}}

\institute{Sternwarte der Universit\"at Bonn, Auf dem H\"ugel 71,
  53121 Bonn, Germany 
  \and
  Max-Planck-Institut f\"ur Astrophysik, Karl-Schwarzschild-Str. 1,
  85740~Garching, Germany}

\maketitle

\begin{abstract}
  Type Ia supernovae\index{Supernova} (SNe) are the best standard
  candles available today in spite of an appreciable intrinsic
  variation of their luminosities at maximum phase, and of probably
  non-uniform progenitors.  For an unbiased use of type Ia SNe as
  distance indicators it is important to know accurately how the
  decline rate and colour at maximum phase correlate with the peak
  brightness.  In order to calibrate the Hubble diagram of type Ia
  SNe, i.e. to derive the Hubble constant, one needs to determine the
  absolute brightness of nearby type Ia SNe.  Globular cluster systems
  of early type Ia host galaxies provide suitable distance indicators.
  We discuss how Ia SNe can be calibrated and explain the method of
  Globular Cluster Luminosity Functions (GCLFs).  At present, the
  distance to the Fornax galaxy cluster is most important for deriving
  the Hubble constant.  Our present data indicate a Hubble constant of
  $H_0=72\pm4$\,km\,s$^{-1}$\,Mpc$^{-1}$.
  
  As an appendix, we summarise what is known about absolute magnitudes
  of Ia's in late-type galaxies.
\end{abstract}

\section{Introduction}

\subsection{Supernova Classification}
\index{Supernova!classification}

When the ``new'' star in Andromeda appeared in 1885 with its striking
association with the Great Andromeda Nebula, it was still a long way
to go until the recognition that these phenomena represent, like
nothing else, our deep relation with the Universe.  Shortly after
1920, astronomers learned that the Andromeda nebula was located far
outside our Milky Way and that the ``Nova'' S Andromedae must have
been much more luminous than any other nova.  Thus Fritz Zwicky
created the term \emph{supernova} (SN) for these events, of which by
1930 only a handful had been detected in external galaxies.

Today it is common wisdom that supernovae are largely responsible for
the enrichment of the universe with heavy elements and thus also
enable the existence of life.  Supernova detections are no longer a
matter of incident alone: Sophisticated supernova searches with
automatic telescopes are now routinely conducted and, thanks to these
efforts, we have detected 1387 SN events until February 1999.

Morphological classification is one of the first scientific steps on
the way to the understanding of a new phenomenon.  It is amazing that
the first attempts to classify SNe still provide the basis, although
our classification scheme has become more subtle and we now know that
the morphological phenomenon ``Supernova'' embraces physically
distinct processes (for the early years see the excellent review of
Trimble \cite{trimble:82}, a recent review on SNe Ia as distance
indicators is from Branch \cite{branch:98}).

As did Minkowski in 1936, we still basically distinguish two types of
SNe, types I and II.  This classification is purely spectroscopic:
type I events do not show hydrogen lines in their spectra, while type
II events do so.  \index{Supernova!parent population} An important
observation can be made with respect to the parent population of SNe:
type I appears in all sorts of galaxies, while type II seems to be
restricted to galaxies with a young stellar population, i.e. spiral
galaxies and irregulars.  For many years, the canonical interpretation
was that the progenitors of type I stem from an old stellar
population.  A popular model is a matter-accreting white dwarf in a
binary system which becomes unstable when the Chandrasekhar mass limit
is reached.  Then the star explodes by thermonuclear detonation or
deflagration \cite{woosley:90} (which is just another expression for
super- or subsonic movement of the burning front but is nevertheless
very important in modelling the event), but see Kundt \cite{kundt:98}
for difficulties of this model.  On the other hand, type II was
thought to be the explosion of a massive star, whose core collapses at
the end of its evolution, forms a neutron star (already Zwicky had
this vision), and the released gravitational energy (about
$10^{53}$\,erg) drives the expansion of the outer shells (in a
complicated and not yet fully understood manner).  Regarding the
energetics of type II events, it soon became clear that radiation and
the kinetic energy of the shell must be only a tiny fraction of the
total energy ($10^{51}$\,erg), the rest being emitted as neutrinos.
The detection of neutrinos from SN\,1987A was a triumph to this theory
(e.g.  \cite{mccray:93}).

The coarse classification depending on the spectral hydrogen features
is further refined.  The subclassification of type II by the shape of
the light curve appears useful, distinguishing SNe II with (II\emph{p}
for \emph{p}lateau) or without plateaus (II\emph{l} for
\emph{l}inear), which basically depends on the thickness of the
hydrogen shell and the speed at which the \element{H}-ionisation front
moves inward during expansion.  But type I SNe are also subclassified
into Ia, Ib and Ic.  By 1983, a few type I SNe had been observed which
developed strong radio emission a few weeks after maximum light.  This
is now commonly interpreted as the core collapse and subsequent
explosion of the outer layers of a massive evolved star (for instance
a Wolf-Rayet star) which lost its hydrogen shell prior to the
explosion.  This new type of SN without hydrogen lines and massive
progenitor was named Ib or Ic, depending on whether helium lines were
strong or not.  The remaining type Ia is spectroscopically distinct
from Ib by the occurrence of \element{Si} lines in an otherwise very
similar spectrum.

For more information on SN classification refer to Wheeler \& Harkness
\cite{wheeler:90} and Filippenko \cite{filippenko:97}.

\subsection{The parent population of type Ia SNe}
\index{Supernova!parent population}

\begin{figure}[htbp]
  \begin{center}
    \includegraphics[bb=42 84 554 787,    
    angle=-90,width=0.7\textwidth,clip]{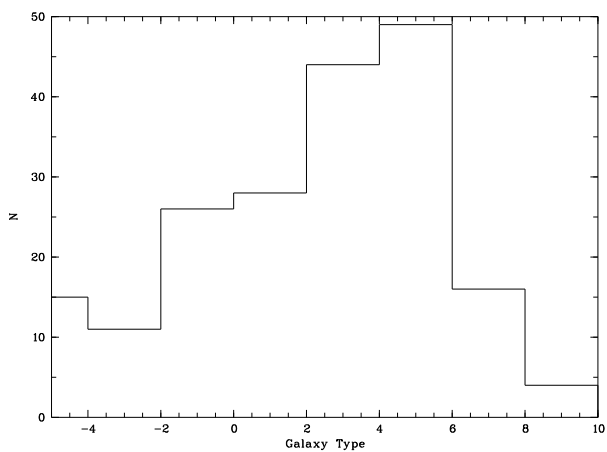}
  \end{center}
  \caption{Distribution of type Ia SNe depending on the host galaxy 
    type (\mbox{de~Vaucouleurs'} classification).  It is apparent that
    Ia events in early-type galaxies are rarer than in late-type
    galaxies.}
  \label{fig:hosts}
\end{figure}

In a modern compilation (Asiago catalogue \cite{barbon:99}), we now
have 403 classified Ia events at the time of writing (May 1999).
Their distribution among the host galaxies is shown in
Fig.~\ref{fig:hosts}.  The abscissa is de~Vaucouleurs' classification,
where the extreme numbers are $-5$ (elliptical galaxies) and $10$
(late-type irregulars).  It is apparent that Ia in elliptical galaxies
are rarer than in spiral galaxies.  If we normalise to luminosity,
then the difference becomes even more obvious.

Table~\ref{tab:SNstat} shows the SN Ia rate in different galaxy types
according to \mbox{Cappellaro} et~al. \cite{cappellaro:93}.  If we
recall that a typical $(M/L)_V$-ratio in elliptical galaxies is about
8 and in spiral galaxies about 0.05, then it becomes clear that a unit
mass in spiral galaxies produces many more SNe than in elliptical
galaxies.  This would be understandable if spiral galaxies had more
opportunities to let Ia's explode than elliptical galaxies have.
Indeed, the former hypothesis, that all Ia's are explosions of
Chandrasekhar mass white dwarfs in old population environments, now
stands on very weak grounds.  Recent investigations have revealed that
there exists a weak statistical correlation with spiral arms in spiral
galaxies \cite{bartunov:94}, which speaks in favour of the
contribution of a parent population of massive stars for at least some
Ia's.  On the other hand, in elliptical galaxies, at best very few
massive stars exist, so one is driven towards the conclusion that Ia's
originate in young \emph{and} old populations.  A further, quite
irritating observation is that Ia's seem to avoid the central regions
of ellipticals and spirals \cite{wang:97}, so that the old bulge
populations apparently are not effective producers of Ia events.  Some
researchers even propose that every supernovae has a massive
progenitor, e.g. Kundt \cite{kundt:90}.

However, our theme is not the physics of Ia's but how they can be
exploited to obtain the Hubble constant.  In the next section, we
shall see that in spite of a probably non-uniform origin, Ia's exhibit
stunningly homogeneous properties which make them excellent standard
candles.
  
\begin{table}
  \caption{SN rates for various galaxy types in\protect\newline
    \mbox{SNu = \# of SNe 
    $\times\left[L_\mathrm{host\,gal.}/(10^{10}L_\odot)\cdot
    t/(100\,\mathrm{yr})\right]^{-1}$}, according to Cappellaro
    et~al. \cite{cappellaro:93}.
    Rates in SNu scale as 
    $\left(H_0/\left(75\,\mathrm{km\,s^{-1}\,Mpc^{-1}}\right)\right)^2$}
  \label{tab:SNstat} 
  \begin{tabular}{c@{\extracolsep{1em}}c} 
    \hline
    Galaxy Type & Ia rate [SNu]\\
    \hline
    E-S0   & $0.13\pm0.06$\\
    S0a-Sb & $0.17\pm0.07$\\
    Sbc-Sd & $0.39\pm0.19$\\ 
    \hline
  \end{tabular}
\end{table}

\section{SNe Ia as distance indicators}

\subsection{The light curve}

\begin{figure}[htbp]
  \begin{center}
    \includegraphics{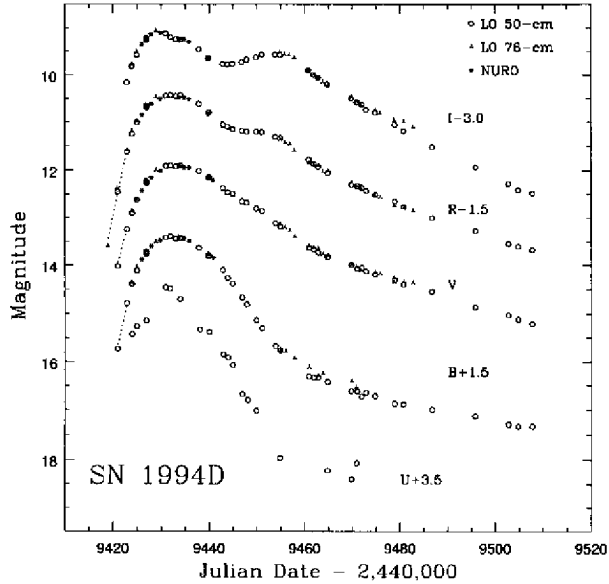}    
  \end{center}
  \caption{Light curves of SN\,1994D \cite{richmond:95} in NGC\,4526
    as an example of typical light-curve shapes of type Ia SNe in the
    optical passbands.}
  \label{fig:lightcurve}
\end{figure}

It has long been noted that the overall shapes of SN Ia light curves
are surprisingly similar.  Figure~\ref{fig:lightcurve} shows a
``typical'' light curve.  After the maximum, a slow decline follows
which has a smooth transition to a linear part.  A linear relation
between time, which enters linearly, and a logarithmic parameter,
magnitude, reflects an exponential law.  The suspicion that the light
curve is triggered by the decay \element[56]{Ni}$\rightarrow$%
\element[56]{Co}$\rightarrow$\element[56]{Fe} was therefore expressed
quite early \cite{colgate:69}.
Meanwhile, beside the case of SN\,1987A \cite{arnett:89}, there is
evidence from the observation of the \element[56]{Co} line in several
supernovae \cite{kuchner:93}.

\subsection{The Hubble diagram}
\index{Hubble diagram}

If one tried to compare the maximum luminosities of SNe by distance
determinations of the respective, necessarily nearby, host galaxies,
the error in the distance would probably dominate the scatter in the
maximum brightness, unless the accuracy of the distance determination
is very high.  A better way of assessing the quality of SNe Ia as
distance indicators is therefore to plot them in a so-called ``Hubble
- diagram'', which uses the apparent peak magnitude as ordinate and
$\log cz$ as abscissa, where $c$ is the speed of light and $z$ the
redshift of the host galaxy.  $cz$ is sometimes called the `recession
velocity' because for small $z$, the cosmological expansion is not
distinguishable from a Doppler effect.

\begin{figure}[htbp]
  \begin{center}    
    \includegraphics[bb=40 85 558
    633,angle=-90,width=0.7\textwidth,clip]{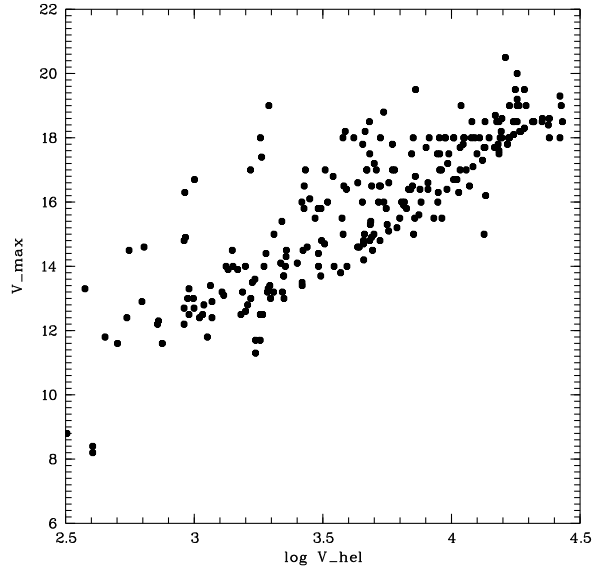}
  \end{center}
  \caption{The Hubble diagram from 197 type Ia SNe taken from the
    Asiago SN catalogue \cite{barbon:99}.  Plotted is the apparent
    maximal V-brightness versus the logarithm of the heliocentric
    recession velocity of the host galaxy.}
  \label{fig:asiago}
\end{figure}

Figure~\ref{fig:asiago} shows the Hubble diagram for 197 classified Ia
SNe, with the heliocentric $cz$ and the visual apparent maximum
magnitude plotted.  From $m-M=5\log r-5$ and the linear approximation
of the redshift-distance relation $cz = H_0\cdot r$ for small $z$, one
obtains
\begin{equation}
  \label{eq:mcz}
  m=5\log cz \underbrace{-5\log H_0+M+25}_{Z}.
\end{equation}
Thus the slope in such a diagram is always 5 while the zero-point $Z$
is the only free parameter, which has to be fixed by observations like
in Fig.~\ref{fig:asiago}.  Throughout this text, we regard $cz$, $H_0$
and $r$ as given in their canonical units, viz. km\,s$^{-1}$,
km\,s$^{-1}$\,Mpc$^{-1}$ and Mpc, respectively.
If all Ia's had the same absolute maximum magnitude $M$, the distance
and thus the brightness of one SN would principally suffice to
determine the Hubble constant:
\begin{equation}
  \label{eq:h0}
  \log H_0=0.2\cdot(M-Z)+5.
\end{equation}
In reality, one would of course aim at determining the distance to as
many SNe as possible.

However, the assumption that all Ia's have the same peak brightness is
at first glance not supported by Fig.~\ref{fig:asiago}.  The
dispersion around the mean relation is more than 1\,mag and there
would be no hope that SNe Ia could be reliable distance indicators
(however, note the sharp borderline of the bright side of the
distribution).  But we must be aware that the sample shown in
Fig.~\ref{fig:asiago} is not a careful selection of good, or even
excellent, observations.  Many SNe have badly determined maxima, we
did not correct for extinction, and most of the SNe have been measured
in the pre-CCD era, while it may be important to take account of the
fact that many SNe are embedded in the irregular structure of their
host galaxies.  This is only possible with CCD observations.  During
the 80s and early 90s, the dispersion in the Hubble diagram of SNe
could be reduced to typically 0.4\,mag, which gave some hope that SNe
Ia are indeed good standard candles.

\begin{figure}[htbp]
  \begin{center}
    \includegraphics[width=1.0\hsize]{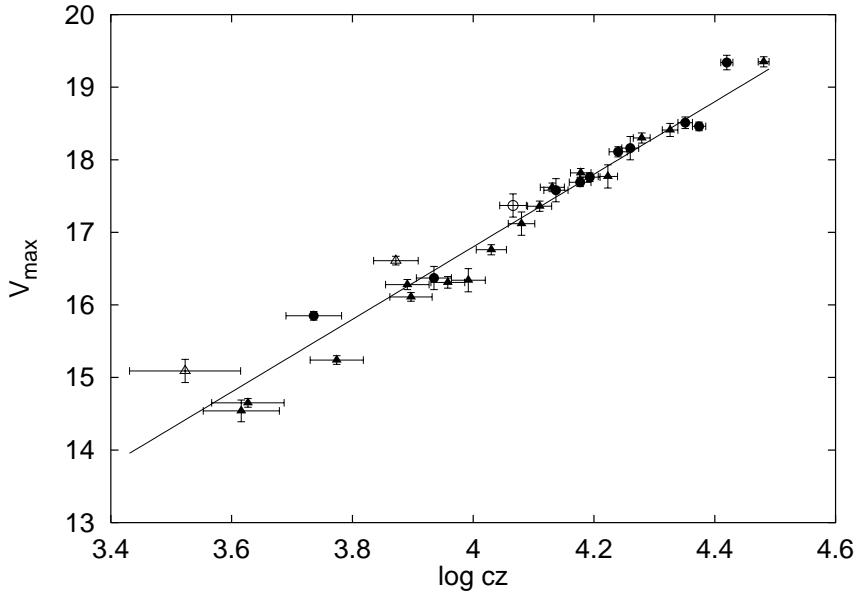}
  \end{center}
  \caption{The Hubble diagram from the high quality data of the 
    Cal\'an/Tololo sample in the $V$ band. Open symbols: red SNe,
    Triangles: late-type host galaxies, Squares: early-type host
    galaxies.}
  \label{fig:hamuy}
\end{figure}

\sloppypar A new level of accuracy was achieved with the publication
of the Cal\'an/Tololo sample by Hamuy et~al. \cite{hamuy:96}, which
contained 29 Ia SNe.  These authors performed a coordinated search
programme for Ia SNe between 1992 and 1994, with the aim to detect SNe
as early as possible, observe good or even excellent light curves in
$B$, $V$ and $I$ and treat the data in a homogeneous manner.
Moreover, they corrected for foreground extinction, applied the
$K$-correction and referred all redshifts to the rest frame as defined
by the microwave background.  Figure~\ref{fig:hamuy} shows as an
example their Hubble diagram in $V$.  The scatter is now only
0.26\,mag, showing the importance of high-quality CCD observations.

But, equally important, it turns out that the scatter around the mean
line is \emph{not erratic} but shows a \emph{systematic} behaviour.
This sounds strange, but, in fact, was already suspected in the late
70s.  The two important parameters are the \emph{decline rate} of a
given SN and its \emph{colour} at maximum light.  Reviving earlier
work, Phillips \cite{phillips:93} again collected evidence for the
fact that Ia SNe with a slow decline after maximum light are
intrinsically brighter than SNe with a fast decline.  To put this in
quantitative terms, the decline rate $\Delta m_{15}$ has been defined
as the decline in magnitudes of the $B$ light curve 15 days after
maximum light.

Regarding the colour, three intrinsically red SNe are marked by open
symbols in Fig.~\ref{fig:hamuy}.  It can be seen that they are
atypical.  They seem to be too faint in comparison with the `normal'
SNe with colours $B_\mathrm{max}-V_\mathrm{max}<0.2$\,mag.  Note that
this red colour and low brightness is not an extinction effect but is
rather an intrinsic feature of these three SNe, as will be made
plausible later on.  The high internal precision of the Cal\'an/Tololo
sample of SNe is a good precondition to investigate further the role
of decline rate and colour.

To model the relation between brightness, decline rate and colour, a
linear approach like
\begin{equation}
  \label{eq:corr}
  m_\mathrm{cor}=m+b\cdot(\Delta m_{15}-1.1)
  +R\cdot(B_\mathrm{max}-V_\mathrm{max})=5\log cz + Z
\end{equation}
is appropriate.  The fit parameters $b$, $R$ and $Z$ must then be
chosen so that the scatter of the corrected magnitudes
$m_\mathrm{cor}$ around a straight line is minimal.
Table~\ref{tab:coef} shows the results of our maximum-likelihood fit
for the Cal\'an/Tololo data in $B$, $V$ and $I$.

\begin{table}[htbp]
  \caption{The fitted decline rate and colour correction coefficients
    and zero points for the Hubble diagram of type Ia SNe based on
    the Cal\'an/Tololo SN sample.  The three red SNe from this sample
    were excluded from the fit to eliminate the influence of any
    peculiar SN events.}
  \label{tab:coef}
  \begin{tabular}{c@{\extracolsep{1em}}ccc}
        & $B$            & $V$            & $I$\\
    \hline
    $b$ & $-0.48\pm0.23$ & $-0.52\pm0.19$ & $-0.40\pm0.22$\\
    $R$ & $-1.51\pm0.62$ & $-0.83\pm0.57$ & $-0.81\pm0.59$\\
    $Z$ & $-3.306\pm0.063$ & $-3.320\pm0.058$ & $-3.041\pm0.060$\\
    \hline
  \end{tabular}
\end{table}

\begin{figure}[htbp]
  \begin{minipage}[t]{0.49\textwidth}
    \includegraphics[width=\textwidth]{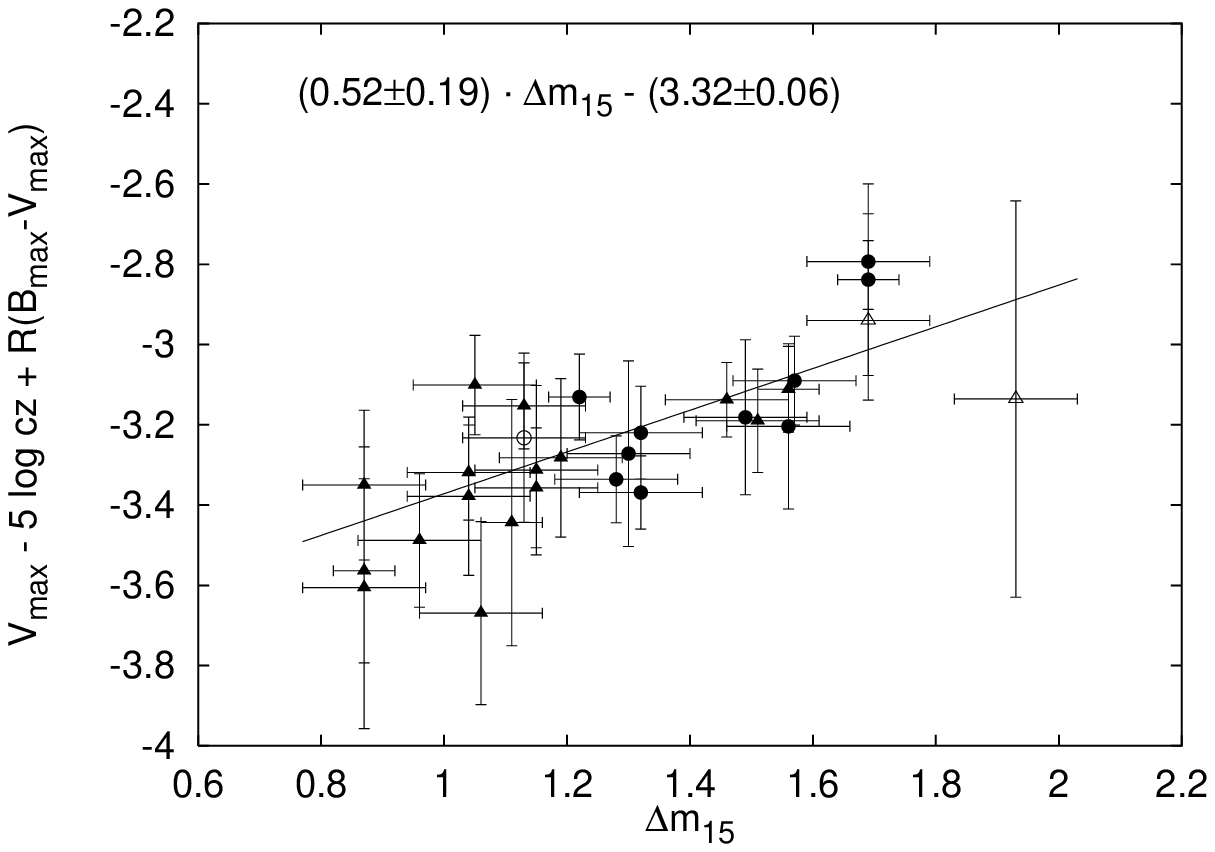}
    \caption{The relation between decline-rate and colour-corrected 
      maximum $V$-magnitude.  The triangles mean SNe in spiral
      galaxies, the circles mean SNe in early-type galaxies. It is
      interesting that in spiral galaxies, all kinds of decline rates
      occur while only fast decliners exist in ellipticals and S0's.
      This could be indicative of the fact that SNe Ia occur in more
      than one population.}
    \label{fig:decline} 
  \end{minipage}
  \hfill
  \begin{minipage}[t]{0.49\textwidth}
    \includegraphics[width=\textwidth]{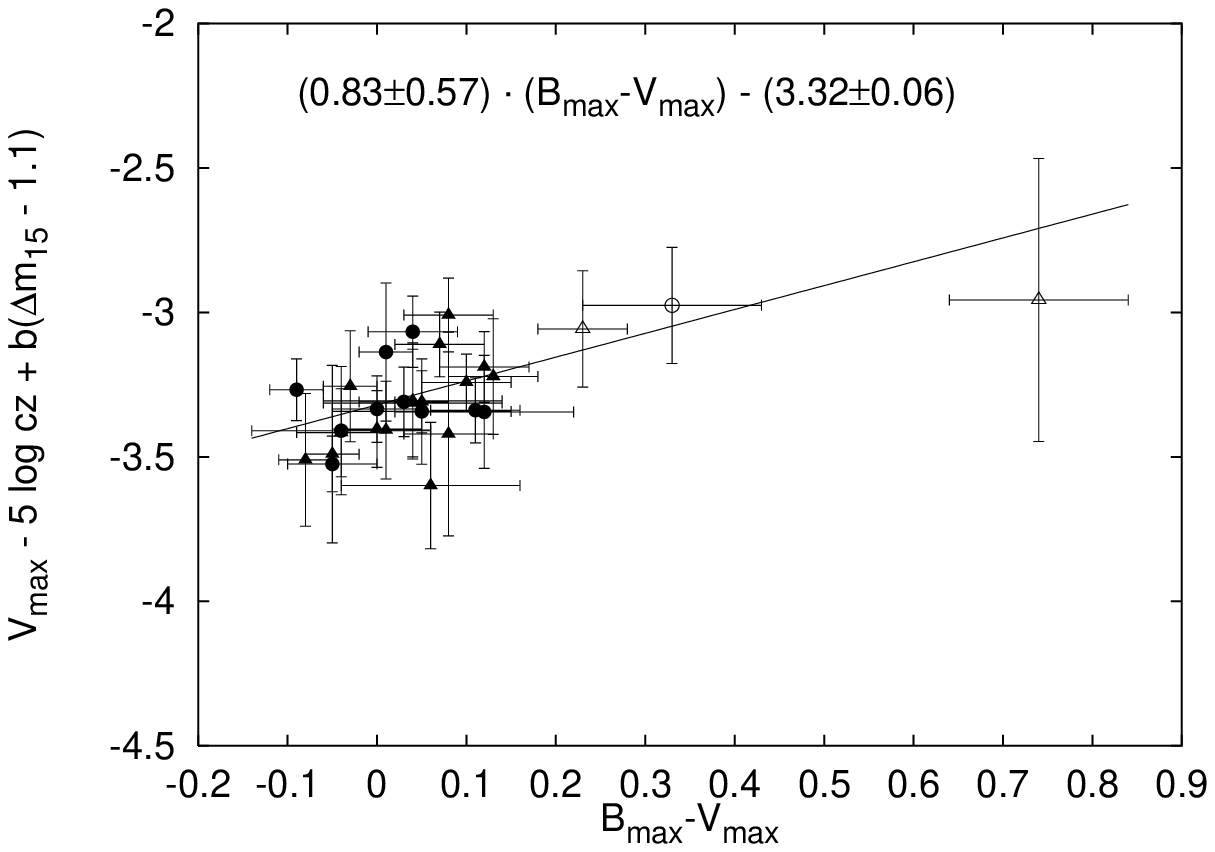}
    \caption{The analogous relation between colour and decline 
      rate-corrected maximum $V$-magnitude.  The fit is done
      \emph{without} the three very red SNe.  One sees that the
      relation for normally coloured SNe also applies to the red
      ones.}
    \label{fig:colour} 
  \end{minipage}
\end{figure}

To uncover the effects of decline rate and colour, we plot the
corresponding magnitude corrections versus decline rate and versus
colour.  Figure~\ref{fig:decline} shows the relation between decline
rate $\Delta m_{15}$ and the colour and redshift-corrected
$V$-magnitude $V_\mathrm{max}-5\log cz
+R\cdot(B_\mathrm{max}-V_\mathrm{max})$.  A correlation, in the sense
that fast decliners are intrinsically fainter, is clearly visible.
Such a behaviour is also theoretically understandable.  If the light
curve is triggered by the decay of \element[56]{Ni}, the available
amount of \element{Ni} should have a considerable effect.  The higher
the \element{Ni} mass, the brighter the SN, and the longer the time
needed for the deposited energy to be radiated away
\cite{hoeflich:96}.  Figure~\ref{fig:colour} shows the relation
between the colour $B_\mathrm{max}-V_\mathrm{max}$, the decline rate,
and the redshift-corrected $V$-magnitude $V_\mathrm{max}-5\log
cz+b\cdot(\Delta m_{15}-1.1)$.  Because most of the SN colours lie in
a small range, the colour correction is smaller than the decline-rate
correction.  As mentioned earlier, the colour correction accounts for
the intrinsic differences of type Ia SNe and not for extinction.
Otherwise the correction coefficient $R=0.83$ for the $V$ band would
have a value similar to $A_V/E(B-V)=3.1$.  But even if the extinction
for a given SN is under- or overestimated, the colour correction
points in the right direction and weakens any extinction error.
  
\begin{figure}[htbp]
  \begin{center}
    \includegraphics[width=1.0\textwidth]{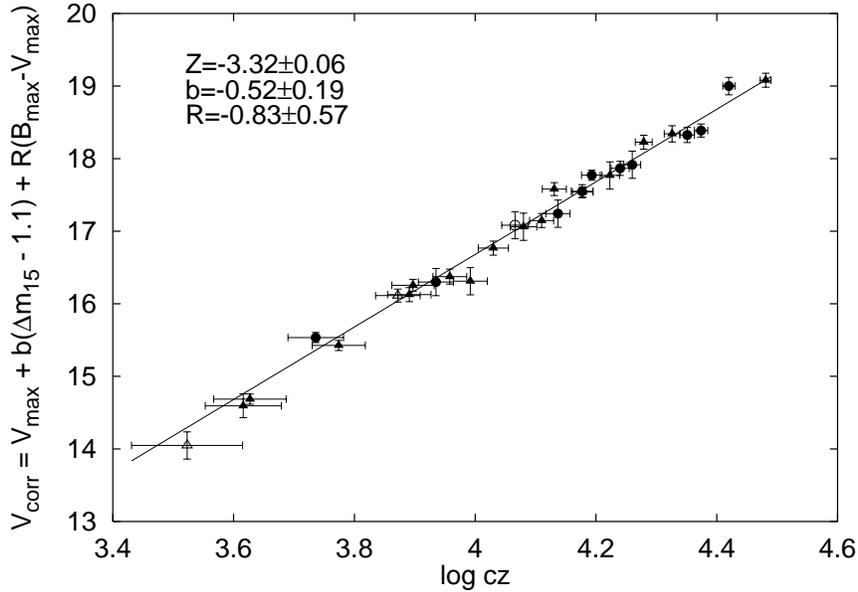}
    \caption{The Hubble diagram of corrected SN magnitudes.  
      The scatter of $\sigma=0.15$\,mag is fully consistent with the
      photometric errors and redshift uncertainties as indicated by
      the error bars, indicating that the \emph{intrinsic} scatter is
      below the measurement accuracy.}
    \label{fig:chdiag}
  \end{center}
\end{figure}

The corrected magnitudes scatter around a straight line only within
the measurement uncertainties.  Type Ia SNe are therefore not perfect
standard candles but almost perfectly \emph{standardisable} candles.
Figure~\ref{fig:chdiag} shows the amazing reduction of the scatter in
in the corrected SN magnitudes in comparison with the diagram of
uncorrected SN magnitudes in Fig.~\ref{fig:hamuy}.

\section{Globular Cluster Luminosity Functions}

The method of distance determination through Globular Cluster
Luminosity Functions has revealed itself as a powerful and accurate
method to derive distances to elliptical and S0 galaxies.  Spiral
galaxies are not suitable, because we need \emph{old} globular
clusters (GCs), which are fewer in number in spirals than in
elliptical galaxies and difficult to detect.
 
\subsection{The principle}

\begin{figure}[htbp]
  \begin{center}
    \includegraphics[width=0.8\textwidth]{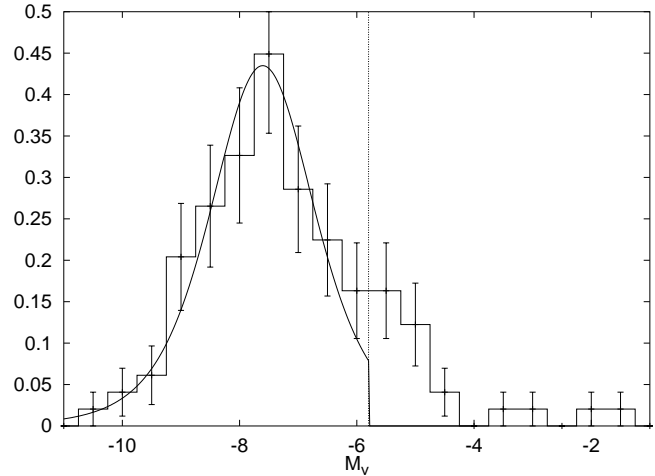}    
  \end{center}
  \caption{The GCLF of the galactic GCS in the $V$ band.  The vertical
    line indicates the magnitude where the distribution becomes
    asymmetric.}
  \label{fig:MW}
\end{figure}

The principle is easily explained: the ``Globular Cluster Luminosity
Function'' (GCLF) is simply the distribution function of the apparent
or absolute magnitudes of globular clusters.  In the literature the
term GCLF is often used also for the histogram of the observed GC
magnitudes related to a given host galaxy.
For example, Fig.~\ref{fig:MW} shows the GCLF of the Galactic system.
We omitted a few clusters due to uncertain extinction corrections.
The peak magnitude in the histogram is called the ``Turn-Over
Magnitude'' (TOM).  The claim now is that this TOM has a
\emph{universal brightness}, which can be calibrated with the Galactic
system and which accordingly can serve as a distance indicator.  It is
important to realise that this TOM does not correspond to any
characteristic mass.  It is rather a consequence of the fact that the
mass function of GCs can be described as $\mathrm{d}N/\mathrm{d}m \sim
m^{-\eta}$, $\mathrm{d}N$ being the number of clusters in the mass
interval $[m,m+\mathrm{d}m]$, with varying exponent $\eta=\eta(m)$.
The logarithmic binning of magnitude then creates a maximum when the
exponent takes the value $\eta =1$, assuming a constant $M/L$ ratio
\cite{mclaughlin:94}.  It is clear that a universal TOM can exist only
among \emph{old} globular cluster systems because the luminosity of
young GCs depends strongly on the exact age.

While theoretically not well understood in detail, the empirical
evidence for a universal TOM is quite strong.  Not only the system of
M31, also elliptical galaxies in the Virgo and Fornax galaxy clusters
have globular cluster systems (GCSs) whose TOMs, within the
observational errors, agree with that of the Milky Way system.
Although not completely clear, the physical cause for the existence of
a universal TOM is probably the combined effect of a universal mass
spectrum of newly formed clusters together with a dynamical erosion of
clusters (e.g. \cite{whitmore:96}).  In this context, it is striking
that the mass spectrum of molecular clouds in the Milky Way has the
same mass spectrum as for instance the mass-rich end of the mass
spectra of several elliptical galaxies.  If, for any reason, the mass
spectrum of molecular clouds is universal and cluster formation inside
molecular clouds occurs in a stochastical way (regarding masses) then
it is plausible for statistical reasons that the mass spectrum of star
clusters reflects the mass spectrum of the parent molecular clouds, if
the probability for a certain cluster mass is approximately
proportional to the parent cloud's mass.  In this context, it is at
least interesting that the mass spectrum of Galactic molecular clouds
is well represented by a power law with $\eta \approx -1.7$
\cite{heithausen:98,heithausen:99}.  This exponent has also been found
for the mass spectrum of globular clusters at high masses (e.g.
\cite{whitmore:96}).

\subsection{The calibration of the TOM}
\label{sec:TOMcal}

If we want to measure the absolute brightness of the TOM, we have to
know the distances to the Galactic globular clusters.  This field has
expanded dramatically during the last years and even for specialists
it has become increasingly difficult to follow the literature in
detail.  Of course, distances to globular clusters are not mysterious;
what makes it complicated is the demand for \emph{accuracy}.

Distances to globular clusters are normally measured by the brightness
of their horizontal branches.  This brightness (we take the mean
brightness of all HB stars in the $V$-band) $M_V$ depends on
metallicity.  We write
\begin{equation}
  \label{eq:metal}
  M_V=\alpha\cdot\left([\mathrm{Fe/H}]+1.6\right)+\beta.
\end{equation}
Without exaggerating too much, one can say that the determination of
the coefficients $\alpha$ and especially $\beta$ is a key to the
distance scale, to globular cluster ages, and consequently also to the
age of the universe (e.g. \cite{vandenberg:96a}).  Because we regard
the entire globular cluster system, an error in $\alpha$ does not
matter much.  If $\alpha$ is overestimated, clusters of high
metallicities will be underestimated in brightness but this will
largely be compensated by low-metallicity clusters.  That means the
error in $M_V$ and thus in the distance is dominated by the zero-point
error in $\beta$.  Therefore, we adopt the most recent value for
$\alpha$ from the literature and calibrate the zero-point $\beta$ with
the aid of RR\,Lyrae stars and horizontal branches of globular
clusters in the Large Magellanic Cloud (LMC), which correspondingly
becomes our \emph{most important distance standard}.

The three most \emph{fundamental} methods of distance measurements
which do not need to be calibrated by other methods and which are not
dependent on any a priori assumptions are: trigonometric parallaxes,
stellar stream parallaxes, and Baade-Wesselink parallaxes of pulsating
stars, preferably Cepheids.  Unfortunately, nature did not provide us
with classical Cepheids in globular clusters.  Regarding trigonometric
parallaxes, the satellite HIPPARCOS was designed to solve all
problems.  However, things turned out to be more complicated, mainly
due to the fact that most interesting targets, for instance Cepheids,
have large parallax errors even in the HIPPARCOS catalogue.  There is
already some controversy on that subject \cite{seggewiss:98} and we
shall not pursue it here.  See Robichon et~al. \cite{robichon:99} for
the most recent results on open cluster distances from HIPPARCOS
measurements.  The second method, using stellar stream parallaxes, is
anyway reasonably applicable only to the Hyades cluster.  To our
knowledge, HIPPARCOS proper motions has yet to be used to derive a
stream parallax to another cluster.

So we are left with Baade-Wesselink parallaxes.  The principle of this
method is to compare the pulsational radial velocity curve of a
pulsating star with its light curve.  In the black-body approximation,
$L\sim R^2\cdot T_\mathrm{eff}^4$ where $L$, $R$, $T_\mathrm{eff}$ are
the absolute bolometric luminosity, the radius, and the effective
temperature of the star, respectively.  If $r$ and $L_\mathrm{app}$
are the distance and apparent luminosity, respectively, we have
$L_\mathrm{app}\cdot r^2\sim L$, and thus $L_\mathrm{app}\sim
T_\mathrm{eff}^4\cdot (R/r)^2$.  But $R/r$ is the angular diameter of
the star and therefore the integration over the radial velocity curve
gives the linear diameter and thus the distance, once
$T_\mathrm{eff}^4$ can be determined.  However, the practical
application bears complications.  Stars are not black-bodies, and the
ability to measure $T_\mathrm{eff}$ by any observable parameter, for
instance a colour index, was in in former times often a question for
the theory of stellar atmospheres.  Meanwhile there is a sufficient
number of interferometrically measured angular diameters of stars that
there is no need to borrow any assumption from theory.  According to
our assessment, a variation of this method, known as the
``Barnes-Evans method'', which is essentially the adaptation for
observational terminology, is the most promising approach today.
Gieren et~al. \cite{gieren:98} applied this method in the infrared to
16 Galactic Cepheids to calibrate the period-luminosity relationship
(PLR).  With this PLR, they derive a distance modulus of $18.46\pm
0.05$\,mag for the LMC.  Infrared data already exist for LMC Cepheids
in order to apply the Barnes-Evans method to these stars directly, but
at the time of writing (February '99) the results are not yet
published.  Work in the optical has been done by Gieren et~al.
\cite{gieren:94}, who applied the Barnes-Evans method to Cepheids in
NGC\,1866, a young globular cluster in the LMC.  The result is in very
good agreement with the result from the PLR analysis.  Admittedly, one
still finds values between $18.2$ and $18.7$\,mag for the LMC distance
modulus in the literature, but the evaluation of these extreme values
is beyond our scope.  However, it is important to keep in mind that
the LMC is the standard for extragalactic distances.  With
$\mu_\mathrm{LMC}=18.46\pm0.06$\,mag, we derive a zero-point for
(\ref{eq:metal}) of $\beta=0.53\pm0.12$, based on the RR\,Lyrae
brightnesses by Walker \cite{walker:92} and the globular cluster
horizontal branches by Suntzeff et~al. \cite{suntzeff:92}.

With this zero-point, we can now determine the absolute brightness of
the TOM of the Galactic system, for which we use the ``McMaster
Catalogue'' of Galactic GCs \cite{harris:96}.  For the analytical
representation, a Gaussian has often been used.  But a close
inspection of the LFs of the Galaxy, M31 and several GCLFs of Virgo
ellipticals \cite{secker:92,secker:93} even reveals that the GCLF is
better described by a $t_5$ function, which is analytically written as
\begin{equation}
  \label{eq:t5}
  t_5(m;m^0,\sigma_t)=\frac{8}{3\sqrt5\pi\sigma_t}
  \left(1+\frac{\left(m-m^0\right)^2}{5\sigma_t^2}\right)^{-3}
\end{equation}
parameterised by the TOM $m^0$ and the width $\sigma_t$.  These
parameters can be fitted to a given sample most precisely by the
maximum-likelihood method \cite{bevington:92}.

\begin{table}
  \caption{The TOMs of the Galactic globular cluster system in
    $B$, $V$ and $I$. The GC sample suffers in $B$ and $I$
    from selection effects in the photometry data.  Therefore the $B$
    and $I$ TOMs are derived from the $V$ TOM and the mean Galactic
    GCS colours $\langle B-V\rangle$ and $\langle V-I\rangle$.}
  \label{tab:galtoms}
  \begin{flushleft}
    \begin{tabular}{c@{\extracolsep{1em}}rrr} 
                 & \multicolumn{1}{c}{$B$} & \multicolumn{1}{c}{$V$}
                 & \multicolumn{1}{c}{$I$}\\
      \hline
      $M^0$      &$-6.85\pm0.10$ &$-7.61\pm0.08$ &$-8.48\pm0.10$\\
      $\sigma_t$ &$1.06\pm0.07$  &$0.92\pm0.07$  &$0.85\pm0.06$\\
      \hline
    \end{tabular}
  \end{flushleft}
\end{table}

\section{The absolute brightness of type Ia supernovae}

Concentrating our view on Ia SNe in early-type galaxies instead of
spirals entails not only advantages.  A disadvantage is that we have
as targets only early-type galaxies, and since Ia events apparently
avoid elliptical and S0 galaxies, the number of targets is relatively
small.  On the other hand, the advantage of early-type galaxies is
their low dust content, whereas in spiral galaxies, the dust
extinction is a notorious nuisance for both Cepheids and SNe.  A
further advantage is that globular clusters are brighter than
Cepheids, making the method reach further in distance.  Moreover,
suitable SNe must be \emph{well observed}.  This means that the
maximum should be covered and that the SN preferentially should be
observed with a CCD to take account of the galaxy background.
However, many SNe in the HST Cepheid programme (see the Appendix) are
\emph{not} well observed and if one includes an object like SN\,1937C
in IC~4182 (which was the first SN with good spectra), one has to face
controversial historical reconstruction of the light curve.
  
In what follows, we distinguish between SNe in the Fornax cluster and
SNe in other elliptical galaxies.

\subsection{Type Ia supernovae in the Fornax cluster} 

During the last decades, the literature on the Virgo cluster as an
extragalactic distance standard has grown to an uncomfortably
complicated level.  We now know that much of the controversy between
the ``long'' and the ``short'' distance scale has to do with the fact
that the Virgo cluster possesses a complicated substructure and an
appreciable depth of $\approx15\%$ of its distance.

The Fornax cluster in the southern hemisphere is situated at a similar
distance as the Virgo cluster, but its properties are quite different,
making it much more appropriate as a distance standard.  First, it is
a compact, evolved galaxy cluster with only a small depth structure.
Secondly, it is dominated by elliptical rather than spiral galaxies,
alleviating problems with extinction.  Thirdly, 3 Ia SNe have been
observed since 1980.  See Table~\ref{tab:SNfornax} for details.  In
particular, SN\,1992A in NGC\,1380 is one of the best ever observed Ia
events.  The two other Fornax SNe both appeared in NGC\,1316, a
peculiar S0 galaxy, perhaps a merger, with a lot of small-scale dust
structure.  But only SN\,1980N is well observed.  We caution that the
maximum phase of SN\,1981D in $V$ is documented by a single
measurement only, which makes the colour quite uncertain.
 
\begin{table}
  \caption{Type Ia SNe in the Fornax galaxy cluster.  The photometry is
    taken from Hamuy et~al. \cite{hamuy:91,hamuy:96}.  The data from
    SN\,1981D result from our own light-curve fitting (small
    differences to the values reported in \cite{richtler:99} result
    from slight differences in the fit procedure and demonstrate the
    lower quality of 81D).}
  \label{tab:SNfornax}
  \begin{flushleft}
    \begin{tabular}{c@{\extracolsep{1em}}cccccc}
      \hline
      SN & host galaxy & $B_\mathrm{max}$ & $V_\mathrm{max}$ & 
      $I_\mathrm{max}$ &
      $\Delta m_{15}$ & $B-V$\\
      \hline
      1992A & NGC\,1380 & $12.57$ & $12.55$ & $12.80$ & $1.47$ & $0.02$\\
      1980N & NGC\,1316 & $12.49$ & $12.44$ & $12.70$ & $1.28$ & $0.05$\\
      1981D & NGC\,1316 & $12.66$ & $12.36$ &  ---    & $1.18$ & $0.30$\\
      \hline      
    \end{tabular}
  \end{flushleft}
\end{table}

We now need the distance to the Fornax cluster in order to convert the
apparent magnitudes in Table \ref{tab:SNfornax} into absolute
magnitudes.  In the next paragraphs, we discuss the Fornax cluster
distance derived by GCLFs and other methods.
 
\subsubsection{Elliptical galaxies in the Fornax cluster}

The first distance of the Fornax cluster based on GCLFs with CCD data
was determined by Kohle et~al. \cite{kohle:96}.  They analysed the
globular cluster systems of 4 ellipticals in Fornax, not including the
S0 host galaxies.  Table~\ref{tab:kohle} lists the apparent magnitudes
of the TOMs together with the values $\Delta V$, $\Delta I$, then
corrections due to metallicity differences.
\begin{table}[htbp]
  \caption{Turn-over magnitudes of galaxies in the Fornax cluster.}
  \label{tab:kohle}
  \begin{tabular}{c@{\extracolsep{1em}}cccc}
    \hline
    galaxy  &  TOM($V$)  & $\Delta V$ & TOM($I$) & $\Delta I$ \\  
    \hline
    NGC\,1374 & $23.52\pm0.14$& $0.22$ & $22.60\pm0.13$ & $0.09$ \\
    NGC\,1379 & $23.68\pm0.28$& $0.40$ & $22.54\pm0.34$ & $0.15$ \\
    NGC\,1427 & $23.78\pm0.21$& $0.12$ & $22.31\pm0.14$ & $0.06$ \\
    NGC\,1399 & $23.90\pm0.08$& $0.12$ & $22.36\pm0.06$ & $0.06$ \\
    \hline
  \end{tabular}
\end{table}

The metallicity correction is quite important because it accounts for
the systematic metallicity difference between early-type galaxies and
our reference, the Galaxy.  This has the following background: as we
have argued, the universal feature behind the luminosity function is
the mass function.  But at a given mass, globular clusters of
different metallicities also have different luminosities, the
metal-poorer ones being brighter.  This is a well known and well
understood property of stellar models and consequently also of stellar
systems.  The Galactic GCS comprises clusters with a wide variety of
metallicities.  Being a spiral galaxy, only a few of its clusters
belong to the metal-rich bulge, the vast majority belong to the
metal-poor halo.  This is different for elliptical galaxies, where the
bulges are the dominating structure components and, consequently, most
GCs of elliptical galaxies are metal-rich.  One should remark that it
is not clear whether ``halos'' of early-type galaxies exist.  Only in
the case of the S0 galaxy NGC\,1380, one of our host galaxies, has it
been possible to distinguish between a ``bulge'' and a ``halo''
population \cite{kissler:97}.  However, it is plausible that the
proportion of metal-rich to metal-poor GCs is in the case of
elliptical galaxies much more in favour of metal-rich objects.  Since
it is not feasible to measure metallicities for individual GCs in
ellipticals, one uses the \emph{mean} colour of a GCS as a measure of
the mean metallicity and calculates a corresponding correction, if the
GCS under investigation has a different metallicity distribution.
This can easily be done by using stellar models.  Ashman \& Zepf
\cite{ashman:95} provide tables where the relevant correction can be
read off.

The application of that correction is however only reasonable if the
TOM can be derived with an adequate accuracy.  For that reason, we
skip the quite uncertain case of NGC\,1379 when calculating the
error-weighted mean distance modulus from Table~\ref{tab:kohle}.
Including the metallicity corrections, we derive a mean distance
modulus of $31.30\pm0.13$\,mag in $V$ and $30.89\pm0.08$\,mag in $I$
for the elliptical galaxies of Kohle et~al. \cite{kohle:96}.  The
value derived from the $I$ data is too small compared to the $V$-data
value and the distance moduli from other methods presented later on.
It is likely that the $I$ images were not deep enough and gave a wrong
TOM.  We therefore do not take this value into account.

Recently, Grillmair and co-authors \cite{grillmair:99} used the Hubble
Space Telescope to obtain luminosity functions in $B$ for NGC\,1399,
NGC\,1404 and NGC\,1316 (which is a special case and will be discussed
shortly).  The resulting distance modulus values (using our Milky Way
TOMs) are $31.41\pm0.11$\,mag ($B$) and $31.34\pm0.11$\,mag ($V$) for
NGC\,1399, and $31.69\pm0.23$\,mag ($B$) for NGC\,1404.  While the
values for NGC\,1399 confirm the TOMs from Table~\ref{tab:kohle}, the
value for NGC\,1404 is also in agreement with earlier work by Richtler
et~al. \cite{richtler:92}.  Although not entirely conclusive, this
would set NGC\,1404 out of the core of the Fornax cluster, which is
suggested also by its strongly deviating radial velocity and its low
specific frequency of globular clusters.  Note, however, that
surface-brightness fluctuations do not confirm the larger distance for
NGC\,1404 \cite{jensen:98}.  This issue remains open.

\subsubsection{NGC\,1380, host galaxy for SN\,1992A}

As already mentioned, SN\,1992A is one of the best ever observed Ia
SNe and therefore enters the zero point determination of the Ia Hubble
diagram with a high weight.  The GCS of NGC\,1380 has been analysed by
Kissler-Patig et~al. \cite{kissler:97} and Della~Valle et~al.
\cite{della:98}.  Interestingly, it turned out that the GCS could be
separated into an elongated metal-rich bulge and a spherical,
metal-poor halo population.

Due to the excellent data quality (very long exposed NTT data in very
good seeing conditions) it was possible to trace the luminosity
function 1\,mag below the TOM, and thus the GCS of NGC\,1380 is the
deepest observed GCS from ground-based data.  Table~\ref{tab:92A}
lists the apparent brightnesses and the widths of the $t_5$-functions
for $B$, $V$, and $I$.  The metallicity corrections are negligible in
this case.

\begin{table}
  \caption{Listed are the apparent TOMs and $\sigma_t$-values of the 
    $t_5$-functions representing the GCS of NGC\,1380 in the bands
    $V$, $B$, and $R$ according to Della~Valle et~al. \cite{della:98}}
  \label{tab:92A}
  \begin{tabular}{c@{\extracolsep{1em}}cc} 
    \hline
    Filter & TOM & $\sigma_t$\\
    \hline
    $B$   &   $24.38\pm0.09$ & $0.89\pm0.10$\\
    $V$   &   $23.69\pm0.11$ & $0.95\pm0.10$\\
    $R$   &   $23.17\pm0.10$ & $0.98\pm0.10$\\
    \hline
  \end{tabular}
\end{table}

\subsubsection{NGC\,1316, host galaxy to SN\,1980N and SN\,1981D}

The fact that NGC\,1316 hosted two Ia's (photometry can be found in
Hamuy et~al. \cite{hamuy:91}), gives it a particular character.
However, it is not the perfect standard it could be.  First, SN\,1981D
is not very well observed, in particular the $V$-peak maximum has a
large error, making the colour and the corresponding corrections
uncertain.  Secondly, NGC\,1316 is not a well behaved elliptical
galaxy, but is classified as a peculiar S0 galaxy, suspected to have
been formed in its present shape in a merger.  It is known that
globular clusters can form in large numbers in galaxy mergers.  If
this event had happened, say, $10^9$ years ago \cite{schweizer:80},
and a lot of globular clusters had been formed, we would expect a TOM
(if any) at distinctly fainter magnitudes than in the case of the
other Fornax ellipticals, because of the many intermediate-age
globular clusters.

In fact, Grillmair et~al. \cite{grillmair:99} in their HST study see
no TOM at all, but an exponential increase of sources beyond the
magnitude where the TOM could be expected.  They suggest they these
sources could be many younger open clusters formed in the merger, but
then apparently not many bright globular clusters could have been
formed.
 
At the time of writing we \cite{gomez:99} are undertaking a study of
this GCS.  Preliminary results are compatible with the results found
by Grillmair et~al. \cite{grillmair:99} (at least down to $V\sim
24$\,mag) in the central region, which we cannot access with
ground-based data.  We found a $V$-TOM of $23.7$\,mag for the entire
LF, which fits well to the other Fornax ellipticals.  But we found
different $V$-TOM's for our inner ($23.7$\,mag) and outer
($24.3$\,mag) regions.  If the TOM's were modified by additional young
clusters formed in the merger, one would have expected the reverse.

It is also curious that the number of clusters is extremely low for
such a bright galaxy.  Regarding the nature of the cluster system,
this galaxy is very interesting, but it is not clear how reliable the
GCLF is as as distance indicator.  Planetary nebulae remain another
possibility to derive a distance (see below).

\subsubsection{Fornax cluster distances from other methods}

The method of GCLFs which we pursue here is of course not the only way
to determine the Fornax cluster distance.  Even if we have no space to
go into details, it is very interesting to have a quick look at other
means and see what they arrive at.
  
\paragraph{Cepheids}

The core of the Fornax cluster does not contain spiral galaxies, in
which Cepheids can be found.  NGC\,1365, a large barred spiral galaxy,
is situated well outside the core.  The HST has been used to find
Cepheids and to derive a distance based on the period-luminosity
relation.  Madore et~al. \cite{madore:98} quote a distance modulus of
$31.35\pm 0.07$\,mag.

\paragraph{Surface brightness fluctuations}

The method of ``surface brightness fluctuations'' is applicable to
early-type galaxies and bulges of spiral galaxies.  It uses the fact
that the pixel-to-pixel variation in a CCD image of an elliptical
galaxy is not simply the photon noise but depends in a characteristic
manner on the surface density of (unresolved) red giants.  With
increasing galaxy distance, an increasing number of individual red
giants are covered by one pixel and the ``surface brightness
fluctuations'' from pixel to pixel become smaller, approaching pure
photon noise.  This effect is particularly pronounced in the infrared
region.

Jensen et~al. \cite{jensen:98} give distances for 6 elliptical
galaxies in the Fornax cluster.  Their average distance modulus is
$31.36$\,mag.  It is interesting that NGC\,1404 fits well with the
other ellipticals, in mild contrast to its GCLF.

\paragraph{Planetary Luminosity Functions} 

Planetary nebulae can easily be identified in early-type galaxies
through narrow-band photometry with filters centred on emission lines.
The assumption that the luminosity function of planetary nebulae
(PNLF) is universal qualifies PNLFs as distance indicators in a
similar manner as GCLFs.

McMillan et~al. \cite{mcmillan:93} quote $31.14\pm0.14$\,mag as the
mean distance modulus for NGC\,1316, NGC\,1399, and NGC\,1404.
However, this is based on an Andromeda distance modulus value about
0.2\,mag smaller than the modern Cepheid value of M31.  It also
supports the assumption that NGC\,1316 (as a Ia host galaxy) is at the
same distance as the Fornax core.  The excellent agreement of the PNLF
distance with the GCLF distance (although numerically a bit
coincidental) is very satisfactory and moreover is a strong point for
the universality of the GCLF.

\subsubsection{Absolute brightness of Fornax SNe}

We now can calculate the absolute brightness of the Fornax supernovae
by applying decline-rate and colour corrections according to
(\ref{eq:corr}) and the values of Table~\ref{tab:SNfornax}.  Using the
distance modulus of $\mu_\mathrm{Fornax}=31.35$\,mag for all SNe in
the Fornax Cluster, we obtain the values listed in
Table~\ref{tab:SNfornaxMv}.

\begin{table}[htbp]
  \caption{The corrected absolute brightnesses of SNe in the Fornax 
    galaxy cluster.}
  \begin{tabular}{ccccc}
    \hline
    SN & Host Galaxy & $M_{B,\mathrm{cor}}$ & $M_{V,\mathrm{cor}}$ & 
    $M_{I,\mathrm{cor}}$ \\
    \hline
    1992A & NGC\,1380& $-18.99\pm0.16$ & $-19.01\pm0.16$ & $-18.69\pm0.16$\\
    1980N & NGC\,1316& $-19.02\pm0.16$ & $-19.04\pm0.16$ & $-18.74\pm0.16$\\
    1981D & NGC\,1316& $-19.20\pm0.23$ & $-19.28\pm0.24$ &      ---       \\
    \hline
  \end{tabular}
  \label{tab:SNfornaxMv}
\end{table}

\subsection{Ia events in early-type galaxies outside the Fornax cluster}

Not many Ia SNe outside the Fornax cluster are currently known for
which we could measure reliable distances via GCLFs.  Either the SN is
badly observed or the host galaxy has not yet been a target regarding
its GCLF, or planetary nebulae, or surface brightness fluctuations
(see the compilation of Della~Valle et~al. \cite{della:98}).
Primarily, this is SN\,1994D in the S0 galaxy NGC\,4526
\cite{richmond:95}, located in the southern extension of the Virgo
cluster.  This supernova is an interesting case because it has been
suspected to violate the decline-rate--luminosity relation by being
too bright in spite of an overall completely normal photometric and
spectroscopic behaviour \cite{richmond:95,sandage:95}.  However, our
GCLF distance to NGC\,4526 \cite{drenkhahn:99} was the first published
individual distance determination for this galaxy, which shows that
NGC\,4526 lies in the foreground of the Virgo cluster.  We derived a
distance modulus of $30.4\pm0.3$\,mag.

Another interesting galaxy is NGC\,4374 which hosted 3 Ia's!
Unfortunately, this is not of great value for us.  SN\,1957B is too
badly observed, as is SN\,1980I (this latter SN is sometimes labelled
``extragalactic'', located almost in the middle between NGC\,4374 and
NGC\,4406.  The third one, SN\,1991bg, is very well observed but is
quite peculiar \cite{hamuy:96}.  It is one of the faintest and reddest
Ia events ever observed and probably cannot enter the Hubble constant
determination currently with any weight.  Moreover, the distance for
NGC\,4374 is not yet conclusive.  A GCLF for NGC\,4374 has been
determined by Ajhar et~al. \cite{ajhar:94}.  They give a TOM in the
$R$-band of $23.58$\,mag.  Using $-8.14$\,mag as the Galactic TOM in
$R$, one obtains $31.72$\,mag as the distance modulus.  With this
distance for NGC\,4374, it is interesting that the corrections for
decline rate and colour shift even this weird SN into a domain where
its corrected brightness is comparable to that of the Fornax SNe.  The
error is huge, admittedly, but this is a consequence of the large and
uncertain colour correction, because the respective fit has been done
over a small colour interval only.  However, it must be cautioned that
the distance value based on planetary nebulae is smaller
\cite{jacoby:90}, leading to a brightness which is definitely too
small to be compatible with other Ia SNe.

Table~\ref{tab:SNother} lists the resulting absolute magnitudes for
SN\,1994D and SN\,1991bg.  The fact that for SN\,1991bg the absolute
magnitudes differ strongly from colour to colour is evidence that the
colour correction must be improved before such a strong extrapolation
can be reasonably done.
\begin{table}
  \caption{Listed are the relevant parameters and absolute corrected 
    brightnesses for SN\,1994D and SN\,1991bg based on the distance
    determinations quoted in the text.}
  \label{tab:SNother}
  \begin{tabular}{c@{\extracolsep{0.5em}}crccc}
    \hline
    SN   & $\Delta m_{15}$ & $B_\mathrm{max}-V_\mathrm{max}$ & 
    $M_{B,\mathrm{cor}}$  & $M_{V,\mathrm{cor}}$ & $M_{I,\mathrm{cor}}$ \\ 
    \hline 
    1994D & $1.31\pm0.08$& $-0.05\pm0.04$& $-18.69\pm0.31$& $-18.69\pm0.31$& $-18.44\pm0.31$\\ 
    1991bg& $1.93\pm0.10$& $0.79\pm0.11$ & $-18.55\pm0.56$& $-18.83\pm0.49$& $-19.18\pm0.50$\\
    \hline
  \end{tabular}
\end{table}

Another ``peculiar'' Ia SN, which, according to its probable distance
of the host galaxy, turns out to be too bright, is SN\,1991T in
NGC\,4527 (e.g. \cite{hamuy:96}).  Although its photometric behaviour
is normal, NGC\,4527 should have a distance modulus of about
$29.90$\,mag to bring SN\,1991T in agreement with other SNe.  A
Cepheid distance to this galaxy would be very desirable.

\section{The Hubble constant and conclusions}

Since there is no standardisation yet either in Cepheid distances,
globular cluster distances, fit procedures etc., one must be aware
that the actually derived Hubble constant is sensitive to all these
individual assumptions and uncertainties.  When calculating a Hubble
constant value from SN data it is necessary to assign weights
according to the observational quality of individual SNe.  To keep
things as easy as possible, we calculated the Hubble constant values
for all SNe (except SN\,1991bg) in each filter and then adopt one
value for each SN as listed in Table~\ref{tab:h0}.  The individual
values cannot be simply averaged because they are not independent.
After properly averaging the adopted values we get a resulting Hubble
constant of $72\pm4$\,km\,s$^{-1}$\,Mpc$^{-1}$.
\begin{table}
  \caption{The Hubble constant values are calculated for 4 SNe
    in each filter according to (\ref{eq:h0}) with the data fom
    Tables~\ref{tab:coef}, \ref{tab:SNfornaxMv}, and
    \ref{tab:SNother}.}.
  \label{tab:h0}
  \begin{tabular}{c@{\extracolsep{1em}}cccc}
    \hline
    SN    & $H_{0,B}$& $H_{0,V}$ & $H_{0,I}$& $H_0$ (adopted)\\
    \hline
    1992A & $73.0\pm5.4$ & $72.8\pm5.4$ & $74.2\pm5.5$ & $73\pm6$ \\
    1980N & $72.0\pm5.3$ & $71.8\pm5.3$ & $72.5\pm5.3$ & $72\pm6$ \\
    1981D & $66.3\pm7.0$ & $64.3\pm7.1$ & ---          & $65\pm7$ \\
    1994D & $83.8\pm11.6$& $84.3\pm12.0$& $83.2\pm11.9$& $84\pm12$\\
    \hline
  \end{tabular}
\end{table}

\section{Appendix: Type Ia SNe in spiral galaxies}

Though this article deals primarily with SNe in early-type galaxies,
the major part of the literature on Ia SNe as distance indicators is
related to spiral galaxies.  First, this is simply due to the fact
that the Ia rate in spirals is larger, but secondly also because of
the attractive possibility of identifying Cepheids with the Hubble
Space Telescope and deriving accurate distances by means of the
period-luminosity relation.  Our approach to the Hubble constant would
thus be incomplete to an intolerable level if we would not briefly
consider Ia's in late-type galaxies and compare them with those in
early-type galaxies.  Rather than discussing the literature, we shall
re-derive distances and absolute supernova magnitudes in the same way
as we have done for SNe in early-type galaxies.  For that we set
readability and didactic argumentation at low priority.  The results
will be squeezed into a few tables, which then have the character of
an appendix.

\subsection{The method of deriving distances}

Some Ia events occurred in nearby spiral galaxies, where the Hubble
Space Telescope could detect and measure Cepheids.

We mainly encounter two difficulties.  For the older SNe, the problem
lies with the photometry of the given SN rather than with the distance
of the host galaxy.  For example, a SN like SN\,1895B in NGC\,5253 is
not suitable because of its uncertain photometry.  Sometimes, the
light curve is more a matter of historical reconstruction than of
exact measurements, for instance in the case of SN\,1937C
\cite{pierce:95}.  We are left with only a handful of promising
targets, see Table~\ref{tab:Ialate}.  But even among those, not all
light curves are convincingly reliable, especially those observed in
the pre-CCD era, because diaphragm photometry does not allow the
subtraction of the local background, which in the case of spiral
galaxies with their small-scale variations is more important than for
early-type galaxies.  The second difficulty is the uncertain
extinction towards the SNe and the Cepheids.  All Cepheid observations
with HST have employed the $V$ and $I$ filters, but only two of all SN
light curves which come here into question are given in $I$, where the
extinction is smallest.

To be consistent in the distance scale, we re-derived all distance
moduli on the basis of published Cepheid photometry.  We cannot
discuss all SNe in detail and refer the reader to the cited
literature.  For each Cepheid in a given galaxy, we compute its
$m-M$-value by applying the P-L-relations~\cite{gieren:98}
\begin{equation}
\label{eq:plr}
\begin{array}{c}
M_V=-2.769\cdot\log P/\mathrm{d}-1.294\\
M_I=-3.041\cdot\log P/\mathrm{d}-1.726
\end{array}
\end{equation}
and take the mean for all Cepheids.  Sometimes, it seems reasonable to
skip individual Cepheids, because of a too striking deviation.  A
respective remark is given in these cases.  We obtain a larger
magnitude difference $m-M$ in $V$ than in $I$ except for IC\,4182.  In
principle, this difference for $V$ and $I$ is interpreted as a
difference in the mean extinction.  We now have to assume a universal
reddening law (which can be erroneous), and calculate the extinction
according to $A_{B,V,I} = R_{B,V,I}\cdot E(V-I)$, where $R_{B,V,I}$
takes the values, $2.56$, $1.93$, $0.93$, respectively
\cite{rieke:85}.  The errors $\Delta(m-M)_V$ and $\Delta(m-M)_I$ are
the dispersions of the individual Cepheid values divided by the square
root of the number of Cepheids.  From $(m-M)_V - A_V = (m-M)_I - A_I$
and the mean reddening $E(V-I)=(m-M)_V-(m-M)_I$ we obtain the distance
modulus $\mu = R_V\cdot (m-M)_I - R_I\cdot (m-M)_V$.  Note that
$R_V-R_I=1$ by definition.

Table~\ref{tab:Ialate} lists those SNe which come into question.
\begin{table}
  \caption{This table lists the suitable Ia events 
    in late-type galaxies where the distance can be derived from
    Cepheids.}
  \label{tab:Ialate} 
  \begin{tabular}{c@{\extracolsep{1em}}ccc} 
    \hline
    SN  & Host galaxy & Ref. for SN & Ref. for Cepheids \\
    \hline
    1937C & IC\,4182  & \cite{pierce:95}   & \cite{saha:94}   \\
    1960F & NGC\,4496 & \cite{schaefer:96} & \cite{saha:96a}  \\
    1972E & NGC\,5253 & \cite{hamuy:96}    & \cite{saha:95}   \\
    1974G & NGC\,4414 & \cite{schaefer:98} & \cite{turner:98} \\   
    1981B & NGC\,4536 & \cite{schaefer:95} & \cite{saha:96b}  \\
    1990N & NGC\,4639 & \cite{hamuy:91}    & \cite{sandage:96}\\
    \hline
  \end{tabular}
\end{table}
Van~den~Bergh \cite{vandenbergh:96b} is a useful reference for a
compilation of most of the data.  We did not include SN\,1998bu in
NGC\,3368 (M96) \cite{suntzeff:99} because of its red colour at
maximum light.  This supernova is clearly reddened but the relative
contributions of reddening and intrinsic colour remain uncertain.

The distance moduli quoted in Table~\ref{tab:cepheids} normally do not
differ much from the values given in the literature.  In a few cases,
differences may amount to about 0.2\,mag, depending on the details of
the Cepheid selection which can influence the adopted extinction.
\begin{table}
  \caption{Listed are the number of Cepheids used for the distance 
    determination, the magnitude difference $m-M$ for $V$ and $I$, the
    resulting reddening (the difference of the distance moduli for
    IC\,4182 is negative!), the extinction-corrected distance moduli,
    and remarks regarding details of the Cepheid selection.}
  \label{tab:cepheids} 
  \begin{tabular}{c@{\extracolsep{0.7em}}rccrcc} 
    \hline
    Host gal. & \# & $(m-M)_V$ & $(m-M)_I$ &
    \multicolumn{1}{c}{$E(V-I)$} & $\mu$ & Remarks \\  
    \hline
    IC\,4182 & 27& $28.21\pm0.05$ & $28.31\pm0.05$ & 
    \multicolumn{1}{c}{0 (def.!)} & $28.26\pm0.07$& \\
    NGC\,4496& 31& $31.06\pm0.05$ & $30.97\pm0.05$ & $0.09\pm0.07$& $30.89\pm0.11$& \\
    NGC\,5253& 8 & $27.99\pm0.07$ & $27.77\pm0.12$ & $0.20\pm0.14$& $27.57\pm0.24$& 1\\
    NGC\,4414& 8 & $31.43\pm0.09$ & $31.36\pm0.08$ & $0.07\pm0.12$& $31.30\pm0.17$& 2\\
    NGC\,4536& 31& $31.07\pm0.05$ & $30.97\pm0.05$ & $0.10\pm0.07$& $30.89\pm0.11$&  \\
    NGC\,4639& 8 & $31.71\pm0.08$ & $31.68\pm0.05$ & $0.03\pm0.09$& $31.65\pm0.12$& 3\\
    \hline
  \end{tabular}\\
  1: largest period skipped. 2: smallest modulus skipped.
  3: largest modulus skipped, perhaps differential reddening, 
  best Cepheids selected
\end{table}

The data in Table~\ref{tab:decline} are the necessary input for
computing the \emph{corrected} absolute magnitudes of the SNe as we
have done for the SNe in early-type galaxies.  Then we calculate the
corresponding values of the Hubble constant according to
(\ref{eq:h0}).  The results are given in Table~\ref{tab:hubble}.

We expect from the Hubble diagram of the Cal\'an/Tololo sample that
the scatter among the corrected magnitudes is of the order 0.1\,mag.
We therefore do not interpret the large scatter of, for example in
$M_B$, 0.9\,mag as being real, but being due to errors.  The distances
are in most cases well determined, one may rather suspect the
photometry of the SNe to be unreliable.  Because we know from the
Cal\'an/Tololo and CfA-sample that SNe in early-type galaxies and
late-type galaxies are indistinguishable after the appropriate
corrections, we presume that SN\,1974G and SN\,1960F are the
problematic cases.  In fact, both are SNe with considerable
photometric uncertainties.  In the case of SN\,1960F, the $V$
magnitude at maximum is not well constrained, which gives much freedom
with respect to the colour correction.

When we skip these two SNe, the Hubble constant (in
km\,s$^{-1}$\,Mpc$^{-1}$) for $B$, $V$ and $I$ results to be
$72.0\pm2.3$, $72.9\pm2.5$, and $73.4\pm6.7$.  We emphasise that these
errors are of ``internal'' nature, i.e. they measure the uncertainty
of the applied method.  The uncertainty resulting from a possible
error in the distance scale is not included.  However, one can regard
all Cepheid distances as being based on a distance modulus of the
Large Magellanic Cloud of $18.50$\,mag.  An increase of 0.1\,mag
approximately results in a decrease of 3\,km\,s$^{-1}$\,Mpc$^{-1}$ of
the Hubble constant value.  Infrared Barnes-Evans distances to LMC
Cepheids will probably give the most accurate values in the near
future.

\begin{table}
  \caption{This table lists the apparent brightnesses (not 
    extinction corrected) of our SNe together with the adopted
    reddening and the decline rate.  See Van~den~Bergh
    \cite{vandenbergh:96b} for a discussion of the relevant
    literature.}
  \label{tab:decline} 
  \begin{tabular}{c@{\extracolsep{0.7em}}ccccc}
    \hline
    SN    & $B_\mathrm{max}$ & $V_\mathrm{max}$ & $I_\mathrm{max}$ & 
    $E(B-V)$ & $\Delta m_{15}$\\
    \hline
    1937C & $8.94\pm0.03$  & $9.00\pm0.03$  & --             & 
    $0.0$         & $0.85\pm0.10$\\
    1960F & $11.77\pm0.07$ & $11.51\pm0.18$ & --             & 
    $0.06\pm0.04$ & $1.06\pm0.08$\\
    1972E & $8.49\pm0.14$  & $8.49\pm0.15$  & $8.80\pm0.19$  & 
    $0.05\pm0.02$ & $0.87\pm0.10$\\
    1974G & $12.48\pm0.05$ & $12.30\pm0.05$ & --             &
    $0.16\pm0.07$ & $1.11\pm0.06$\\
    1981B & $12.04\pm0.04$ & $12.00\pm0.07$ & --             & 
    $0.0$         & $1.10\pm0.05$\\
    1990N & $12.74\pm0.03$ & $12.72\pm0.03$ & $12.95\pm0.05$ & 
    $0.0$         & $1.07\pm0.05$\\
    \hline
  \end{tabular}
\end{table}
\begin{table}
  \caption{Given are the absolute \emph{corrected} magnitudes, 
    and the corresponding Hubble constant values as derived with
    (\ref{eq:h0}) and the coefficients from Table~\ref{tab:coef}.}
  \label{tab:hubble} 
  \begin{tabular}{c@{\extracolsep{1em}}ccc}
    \hline
    SN & $M_{B,\mathrm{cor}}$ & $M_{V,\mathrm{cor}}$ & $M_{I,\mathrm{cor}}$ \\ 
    \hline
    1937C& $-19.11\pm0.10$ & $-19.08\pm0.11$ & --\\
    1960F& $-19.61\pm0.31$ & $-19.68\pm0.36$ & --\\ 
    1972E& $-19.07\pm0.35$ & $-19.05\pm0.38$ & $-18.72\pm0.30$\\
    1974G& $-19.41\pm0.23$ & $-19.43\pm0.23$ & --\\
    1981B& $-18.91\pm0.16$ & $-18.92\pm0.17$ & --\\
    1990N& $-18.93\pm0.12$ & $-18.93\pm0.14$ & $-18.70\pm0.25$\\
    \hline
    SN &   $H_{0,B}$    & $H_{0,V}$    & $H_{0,I}$ \\ 
    \hline
    1937C& $69.1\pm3.2$ & $70.5\pm3.6$ & --\\
    1960F& $54.9\pm7.8$ & $53.5\pm8.9$ & --\\ 
    1972E& $70.3\pm11.3$& $71.4\pm12.5$& $73.0\pm10.5$\\
    1974G& $60.1\pm6.4$ & $60.0\pm6.3$ & --\\
    1981B& $75.7\pm5.6$ & $75.9\pm5.9$ & --\\
    1990N& $75.0\pm4.1$ & $75.5\pm4.9$ & $73.7\pm8.6$\\
    \hline
  \end{tabular}
\end{table}

\section{Acknowledgements}

We thank Wilhelm~Seggewiss for many useful discussions and the
permission to use his literature compilation.  We also thank
Gordon~Ogilvie for manuscript reading and Bruno Leibundgut for helpful
and improving comments.
\end{document}